\documentclass[twoside,leqno,twocolumn]{article}  
\usepackage{ltexpprt} 

\usepackage[utf8]{inputenc}
\usepackage{amsmath}
\usepackage{amssymb}
\usepackage{algorithm}
\usepackage[noend]{algpseudocode}
\usepackage{graphicx}
\usepackage{numprint}
\usepackage{color}
\usepackage{eurosym}
\usepackage{url}
\usepackage{paralist}
\usepackage{xspace}
\usepackage{bm}
\usepackage{paralist}
\usepackage{bbding}
\usepackage{hyperref}

\usepackage{tikz}
\usetikzlibrary{bayesnet}

\usepackage[caption=false,font=normalsize,labelfont=sf,textfont=sf]{subfig}

\algdef{SE}[SUBALG]{Indent}{EndIndent}{}{\algorithmicend\ }%
\algtext*{Indent}
\algtext*{EndIndent}

\hypersetup{pdfauthor={Miguel Araujo},pdftitle={BreachRadar}}

\newtheorem{problem}{Problem}

\newcommand{\bit}{\begin{compactitem}}
\newcommand{\eit}{\end{compactitem}}
\newcommand{\ben}{\begin{compactenum}}
\newcommand{\een}{\end{compactenum}}

\newcommand{\methodsc}{{\scshape{BreachRadar}}\xspace}
\newcommand{\method}{BreachRadar}

\begin{document}

\title{\method: Automatic Detection of Points-of-Compromise}

\author{Miguel Araujo\thanks{Carnegie Mellon University and INESC-TEC. \newline\small{maraujo@cs.cmu.edu} .} \\
\and 
Miguel Almeida\thanks{Feedzai. \newline \small{miguel.almeida@feedzai.com, jaime.ferreira@feedzai.com, luis.silva@feedzai.com, pedro.bizarro@feedzai.com}}
\and
Jaime Ferreira\footnotemark[2]
\and
Luis Silva\footnotemark[2]
\and
Pedro Bizarro\footnotemark[2]
}
\date{}

\maketitle


\begin{abstract} \small\baselineskip=9pt 
Bank transaction fraud results in over \$13B annual losses for banks, merchants, and card holders worldwide. Much of this fraud starts with a Point-of-Compromise (a data breach or a "skimming" operation) where credit and debit card digital information is stolen, resold, and later used to perform fraud.
We introduce this problem and present an automatic Points-of-Compromise (POC) detection procedure. \methodsc is a distributed alternating algorithm that assigns a probability of being compromised to the different possible locations. We implement this method using Apache Spark and show its linear scalability in the number of machines and transactions.
\methodsc is applied to two datasets with \textit{billions} of real transaction records and fraud labels where we provide multiple examples of real Points-of-Compromise we are able to detect. We further show the effectiveness of our method when injecting Points-of-Compromise in one of these datasets, simultaneously achieving over 90\% precision and recall when only 10\% of the cards have been victims of fraud.
\end{abstract}

\section{Introduction}

FICO estimates European losses to fraud at \EUR{1.6 billion} in 2014~\cite{fico2015} and global fraud in 2013 was estimated at \$13.9 billion~\cite{ftpartners2015}. If the cost of lost merchandise as well as redistribution costs are included, fraud is estimated to account for 1.32\% of the revenue for the average merchant in the US and it's higher if the merchant operates globally~\cite{lexisnexis201509}. 

A Point-of-Compromise (POC) is a (physical or virtual) location of the payment network, such as an ATM or a point-of-sales terminal, that processed or collected payment information and that was compromised by fraudsters. In a classical scenario, the victim's card is swiped in a small rogue device (possibly installed in an ATM or vending machine, or used by malicious employees whenever the card leaves the owners' sight at a bar, restaurant or gas station) that reads and stores the magnetic stripe data which is then, e.g., sold and written on a new cloned card. However, this is not the only scenario: data breaches are also common POCs which might occur at the merchant or even payment-processor level.

Given the increase in both the number of data breaches and in the number of cards affected (Target's 2013 data breach alone exposed an estimate of 40 million cards~\cite{targetKrebs}), early and accurate detection of POCs is not only vital for fraud prevention, but could also lead to a decrease in the expected loss from these breaches, reducing their frequency. The timely discovery of a Point-of-Compromise could prevent the fraudulent use of other cards compromised at the same location and early detection could prevent thousands of fraud cases. 

\begin{figure*}[ht] 
  \centering 
  \subfloat[\textbf{Significant savings,} over a period of 6 weeks (red), even assuming a \$10 reissue cost per card (green).]{\includegraphics[width=0.31\textwidth, clip, trim = 5mm 0mm 5mm 0mm]{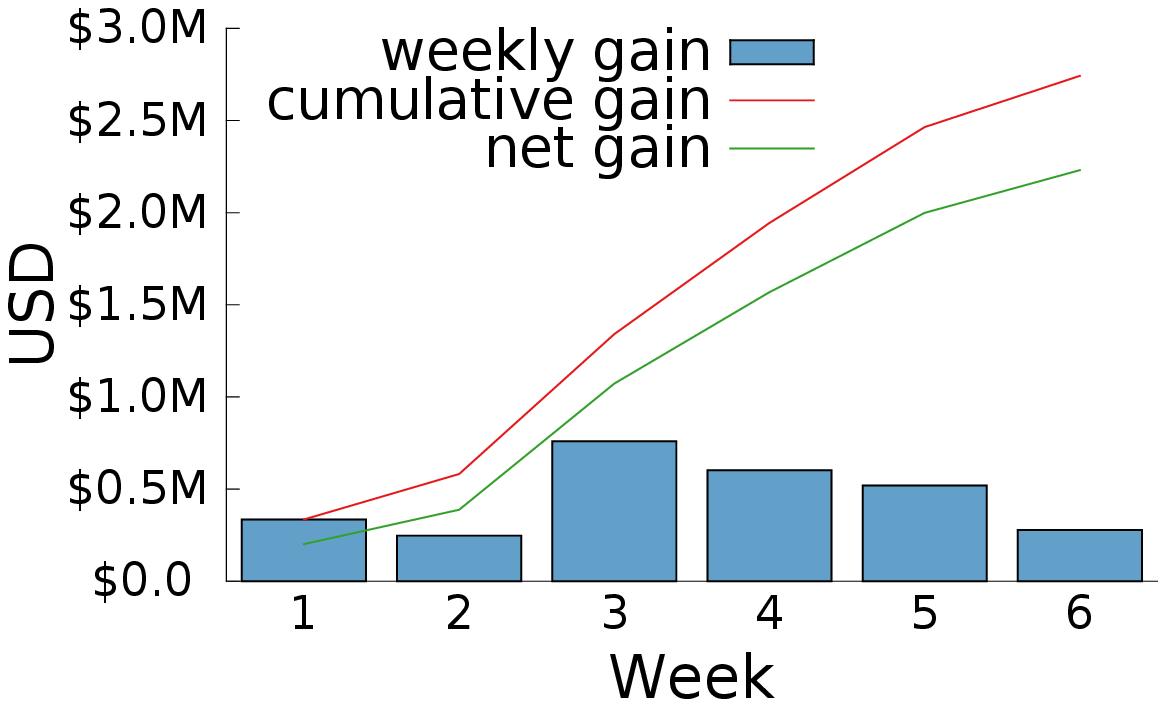} \label{fig:gain}}
  \quad
  \subfloat[Almost perfect \textbf{precision and recall, significantly higher than competing methods.}]{\includegraphics[width=0.31\textwidth, clip, trim = 2mm 0mm 5mm 0mm]{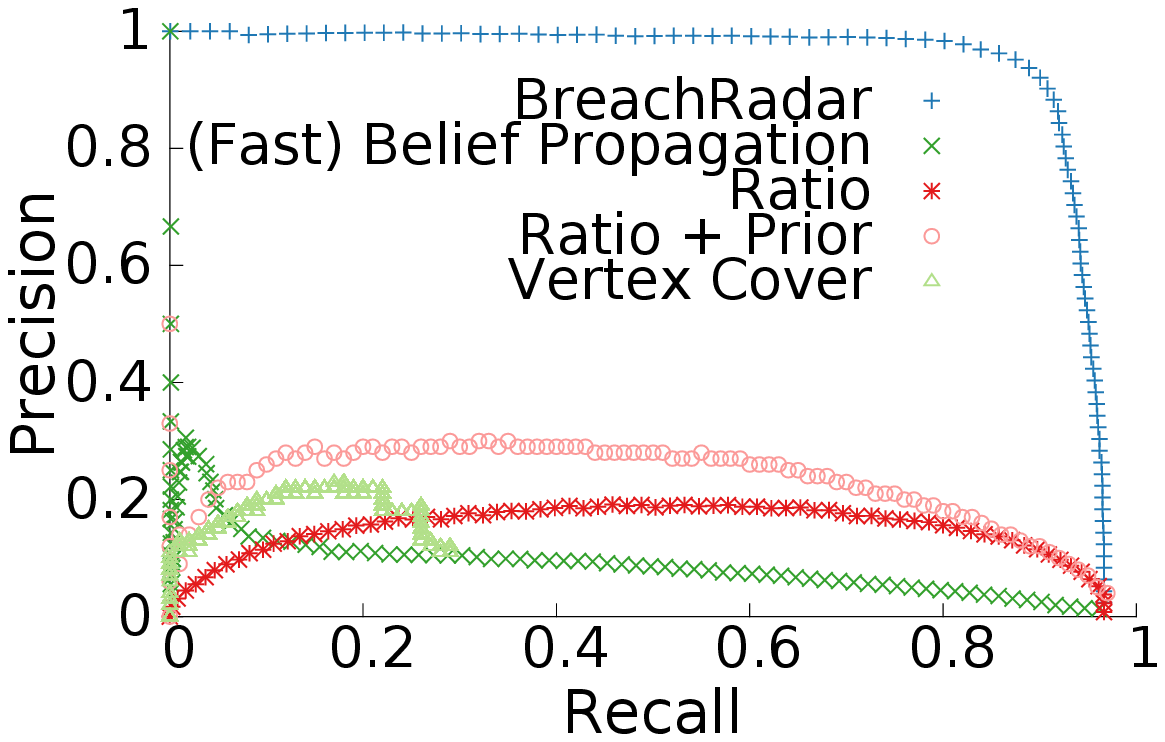} \label{fig:fabppvr}}  
  \quad
  \subfloat[\textbf{Linear speed-up} with number of possible Points-of-Compromise.]{\includegraphics[width=0.31\textwidth, clip, trim = 1mm 0mm 5mm 0mm]{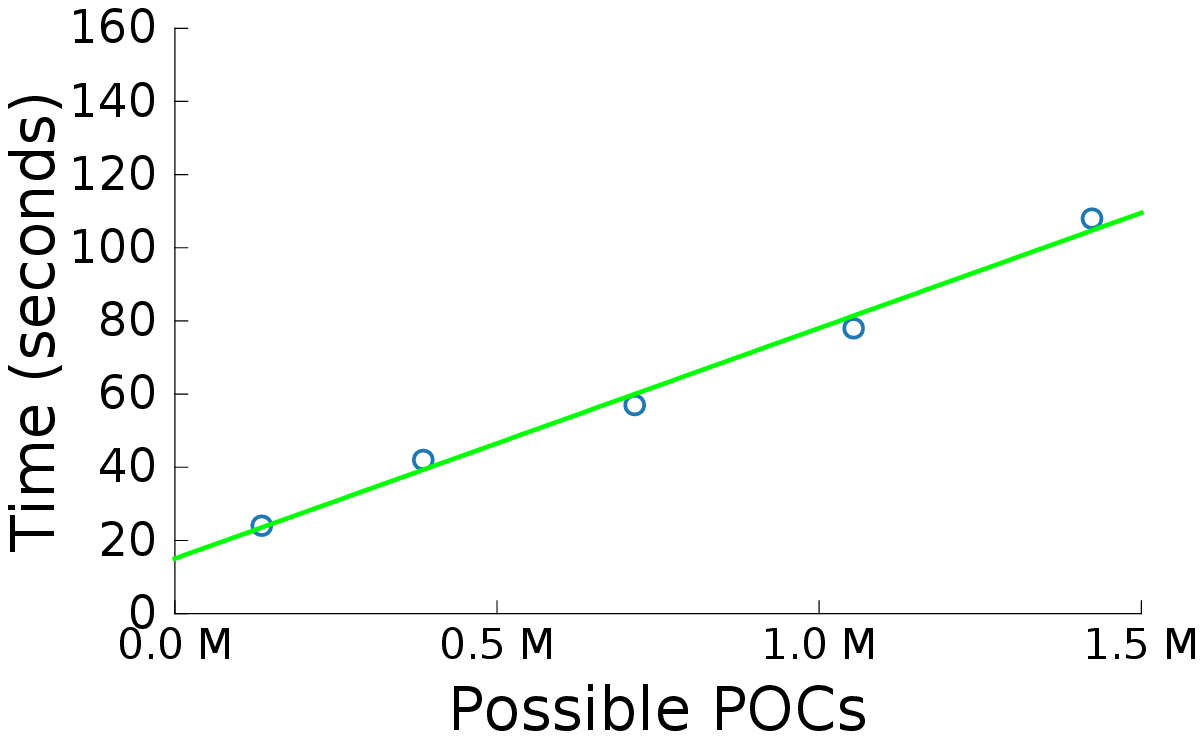} \label{fig:edges_time}}
  \caption{\methodsc is \textbf{fast and effective}.}
\end{figure*}

As an example, Figure \ref{fig:gain} illustrates the weekly savings when \methodsc is applied to one of the two real datasets we explore. By automatically canceling cards that were used in locations marked as potentially compromised, and even after assuming a \$10 reissue cost per card, our system is able to prevent over \$2 million USD in credit-card fraud in a period of just 6 weeks.

The main contributions of this paper can be summarized as follows:

\ben
 \item \textbf{Point-of-Compromise Problem.} We formulate a novel and important POC detection problem. 
 \item \textbf{Effectiveness.} \methodsc accurately identifies \textit{Points-of-Compromise}, achieving over 90\% precision and recall when only 10\% of the stolen cards have been used in fraud (Figure \ref{fig:fabppvr}).
 \item \textbf{Distributed POC-Detection algorithm.} We provide a scalable distributed algorithm for POC detection in big datasets (Figure \ref{fig:edges_time}).
\een

\textbf{Further Applications.} While we focus on the identification of \textit{Points-of-Compromise} in bank transactions, there are other domains where \methodsc could help identify malicious or abnormal activity. We invite the reader to consider any situation where individual nodes might trigger abnormal behavior in their neighbors. Consider anti-virus applications and machine-file bipartite graphs: given malware symptoms in some of the machines, a small set of files that exist in common could be formulated as \textit{Point-of-Compromise} detection problem. Similarly, quick identification of food-poisoning sources or of faulty parts in the car industry can be formulated under this setting.

\section{Background and Related Work}\label{sec:backrw}

While apparently a simple problem, several reasons compound to make the automatic detection of POCs a challenge to naive approaches: a) the variety of \textit{Points-of-Compromise}, e.g., database breaches, card skimming devices, etc; b) the variety of time granularities, e.g. database breaches compromise months of transactions, while an employee skimming cards might do it for a single day; c) the lack of ground-truth labels, as fraudulent charge reports do not identify the origin of the compromise and d) the scale of the problem, as datasets with billions of transactions with millions of possible \textit{Points-of-Compromise} are common.

\subsection{Summary}
Table \ref{tab:salesman} characterizes the most relevant methods described in this section. 
We analyze a method's ability to \textit{find Points-of-Compromise} and to \textit{scale} at least quasilinearly with the number of transactions that need to be processed. We consider that a method has proper \textit{risk assessment} if it doesn't believe that more transactions to safe merchants reduce the probability that the card might have been stolen at a single compromised location. We consider methods to be \textit{Blame-aware} if they acknowledge that cards are likely stolen only once, so they should not significantly contribute to an increased POC likelihood of multiple locations. We consider a method to be \textit{Confidence-aware} if it incorporates the idea that more evidence improves the confidence of a POC label\footnote{As an example, we are more confident that a location has been compromised if 200 out of 600 cards who transacted there were victims, than if 3 out of 6 were.}. Finally, a method is capable of \textit{Early Detection} if it shows high recall even when only a small percentage (e.g., 10\%) of the cards at a location are \textit{fraud-cards} (a card with at least one fraudulent transaction).

\begin{table*}[t]
  \centering
  \caption{Comparison of \methodsc with other methods. Properties are described in Section \ref{sec:backrw}.}
  \label{tab:salesman}
  \begin{tabular}{|l||c||c|c|c|c|c|}
  \hline
     & \footnotesize\textbf{\method} & \footnotesize{Ratio} & \footnotesize{Ratio + Prior} & \footnotesize{Belief Propagation} & \footnotesize{Vertex Cover} & \footnotesize{Real-time Detection} \\\hline
    \footnotesize{Finds POCs} & \Checkmark & \Checkmark & \Checkmark & \Checkmark & \Checkmark &  \\\hline
    \footnotesize{Scalability} & \Checkmark & \Checkmark & \Checkmark & \Checkmark &  & \Checkmark \\\hline
    \footnotesize{Risk assessment} & \Checkmark & \Checkmark & \Checkmark &  & \Checkmark &  \\\hline
    \footnotesize{Blame-aware} & \Checkmark &  &  &  & \Checkmark &  \\\hline
    \footnotesize{Confidence-aware} & \Checkmark &  & \Checkmark &  &  & \\\hline
    \footnotesize{Early Detection} & \Checkmark &  &  &  &  &  \\\hline
  \end{tabular}
\end{table*}

\subsection{Real-time Fraud Detection}
While not able to identify \textit{Points-of-Compromise}, state-of-the-art fraud detection solutions merge statistical, machine learning and data mining tools in order to create models that estimate the fraud probability of individual transactions in real-time. 
For further information on this orthogonal problem, we refer the reader to specific literature~\cite{bolton2002statistical, dal2014learned}.

\subsection{Points-of-Compromise}
Simple metrics are unable to provide an appropriate measure for the likelihood of a point being the origin of a compromise. Ranking locations by the number of \textit{fraud-cards} with which they interacted does not work, as many merchants process many transactions and thus interact with a high number of \textit{fraud-cards}. The ratio of \textit{fraud-cards} shouldn't be used either, as the majority of the locations have small numbers of transactions and high ratios (by chance) do not imply a compromised location.

Current systems for \textit{Point-of-Compromise} detection are typically hindered by these issues. Absolute number of \textit{fraud-cards} and \textit{fraud-card} ratios are commonly used~\cite{patentUS7580891B2, patentUS8473415B2}, perhaps coupled with time-windows~\cite{patentUS8600872B1} to restrict the set of transactions considered. Arbitrary thresholds that indicate whether a merchant was compromised need to be defined, but suggestions have been made that supervised classification algorithms could also be trained, after information about which merchants were in fact compromised is obtained~\cite{patentUS20050055373A1}. A different approach suggests comparing recent fraud-rates at each merchant with their historical fraud-rate and flagging outlier deviations~\cite{patentUS7761379B2}.

\subsection{Guilt-by-association}

Semi-supervised learning techniques, in particular graph-based methods such as Label Propagation, could be used to label nodes as \textit{compromised} or not-\textit{compromised} in the network; labeled nodes \textit{influence} neighbors according to an Homophily Matrix which establishes whether nearby nodes tend to have similar or opposite labels. Some variations extend Label Propagation to incorporate label confidence and prior information, which could be used when only positive labels (fraud) are observed~\cite{yamaguchi2015socnl}. 


However, both methods have the underlying assumption that more connections to innocent nodes imply a smaller likelihood that the node has been compromised; this idea is at the core of guilt-by-association algorithms: similarly to how connections to fraudulent nodes increase the probability of a node being fraudulent, then connections to safe locations decrease this probability. For this problem, we know that the opposite is true: connections to innocent merchants do not compensate for the fact that a connection to a compromised location exists. Our results showing Belief Propagation's low performance corroborate this intuition.


\subsection{Vertex Cover}
This problem can also be formulated as a vertex cover problem in bipartite graphs: given a set of cards who were victim of fraud, we would like to identify the smallest subset $\mathcal{S}$ of adjacent nodes (i.e., merchants) so that every card who has been a victim has at least one adjacent location in $\mathcal{S}$. Unfortunately, this formulation is NP-hard\footnote{This can be easily shown through reduction to Minimum Set Cover, one of Karp's 21 NP-complete problems~\cite{karp1972reducibility}.}. Nevertheless, in Section \ref{sec:comparison} we evaluate a greedy approximation: on each iteration, we consider as compromised the location with the highest number of adjacent \textit{fraud-cards}, remove them from the bipartite graph and repeat.

\section{The POC Problem}\label{sec:problem}
We assume the point-of-view of a payment network or card issuer who has visibility over the majority of the transactions of a set of cards, some of which have been identified as \textit{fraud-cards} (these are typically canceled and reissued). Loosely speaking, our goal is to automatically identify a small set of locations that many \textit{fraud-cards} have in common.

We represent the set of transactions as a bipartite graph with cards on one side and \textit{possible Points-of-Compromise} on the other.

A \textit{possible Point-of-Compromise} is a feature that a subset of transactions have in common, such as a specific point-of-sale terminal, a store identifier or a merchant name (i.e. all the stores from a corporation). In practice, we would also like to incorporate time as a feature as data breaches and skimming devices temporally correlate transactions: in Section \ref{sec:results}, we consider \textit{terminal-week} pairs as \textit{possible Points-of-Compromise}, but many other options (or combinations thereof) are admissible. For simplicity, we use the terms \textit{possible Point-of-Compromise} and location interchangeably. Edges connect two nodes if there is a transaction between a given card and a given location.

%

\subsection{Problem Definition}
Let's consider $\mathcal{C}$ to be the set of all cards and $\mathcal{L}$ to be the set of all locations. $\mathcal{G} = (\mathcal{C} \cup \mathcal{L}, E)$ is the bipartite graph and for every edge $(i,j) \in E \Rightarrow i \in \mathcal{C}$ and $j \in \mathcal{L}$. 
 
We will always use index $i$ to represent cards and index $j$ to represent locations. $\mathbf{L}_i$ is the set of neighboring locations of card $i$ and $\mathbf{N}_j$ is the set of neighboring cards of location $j$.

$f : \mathcal{C} \rightarrow \{0, 1\}$, part of our input, is a function indicating whether a given card $c \in \mathcal{C}$ was a victim of fraud or not.
 
$\mathbf{B}$ is a $|\mathcal{C}|\times|\mathcal{L}|$ matrix where $b_{ij}$ is the probability that $j$ is the location responsible for $i$ being a \textit{fraud-card}, or the blame which card $i$ attributes to possible POC $j$.

Using this notation (summarized in Table \ref{tab:notation}), the POC detection problem can be formulated as:
 
\begin{problem} \textbf{(Spotting Points-of-Compromise)}
 \bit
  \item \textbf{Given:} A graph $\mathcal{G} = (\mathcal{C} \cup \mathcal{L}, E)$ and fraud labels $f : \mathcal{C} \rightarrow \{0, 1\}$.
  \item \textbf{Find:} 
  \bit
    \item $\bm{\theta} : \mathcal{L} \rightarrow [0, 1]$, where $\theta_j$ is the probability that location $j$ is a \textit{Point-of-Compromise};
    \item $\mathbf{B} : \mathcal{C} \times \mathcal{L} \rightarrow [0, 1]$, where $b_{ij}$ is the blame that card $i$ assigns to location $j$.
  \eit
 \eit
\end{problem}
 
 \begin{table}
  \centering
  \caption{Notation, symbols and definitions}
  \label{tab:notation}
  \begin{tabular}{|c|l|} 
  \hline
  \textbf{Symbol} & \textbf{Definition} \\\hline \hline
  $\mathcal{C}$ & Set of all cards \\ \hline
  $\mathcal{L}$ & Set of locations (possible POCs)\\ \hline
  $\mathcal{G}$ & Graph of cards and locations\\ \hline
  $\mathbf{L}_i$ & Set of locations neighboring card $i$\\ \hline
  $\mathbf{N}_j$ & Set of cards neighboring location $j$\\ \hline
  $f_i$ & Boolean indicating if $i$ is a fraud-card\\\hline
  $\bm{\theta}$ & Vector of POC probabilities.\\ \hline
  $\mathbf{B}$ & Blame matrix \\ \hline
  $\mathbf{b_i}$ & A row of matrix $\mathbf{B}$\\ \hline
  $b_{ij}$ & A cell of matrix $\mathbf{B}$\\ \hline
  \end{tabular}
 \end{table}

\section{POC-detection Algorithm} \label{sec:method}
We describe a novel algorithm for the identification of \textit{Points-of-Compromise} following a simple Bayesian inference approach. We adopt the principle that predictions are inherently uncertain and require more evidence in order to increase their confidence, and assume that cards are compromised at a single location and should influence each other towards agreeing on the (locally) most likely mutual \textit{Point(s)-of-Compromise}. 
 
\subsection{A POC Hierarchical Model}\label{sec:model}
We start by assuming that whether a location has been compromised is represented by $p_j$, a Bernoulli random variable whose success probability $\theta_j$ is taken from a Beta distribution (its conjugate prior). 

From the card perspective, we assume that each \textit{fraud-card} has an associated variable $r_i$ taken from a categorical distribution of size $|L_i|$ and probability vector $\mathbf{b_i}$, where each element $b_{ij}$ of $\mathbf{b_i}$ is linearly proportional to the respective $\theta_j$ compromise probability. This means we are implicitly assuming that the probability of a card blaming a location is linearly proportional to the probability of it being a POC.

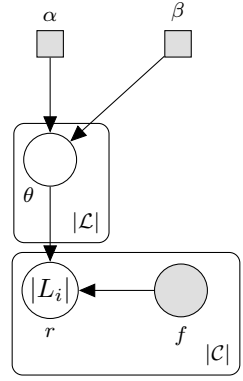
\begin{figure}
  \centering 
  \begin{tikzpicture}

  \node[const, xshift=0cm, label=above:{$\alpha$}] (alpha) {} ;  
  \node[const, xshift=+1.7cm, label=above:{$\beta$}] (beta) {} ;

  \node[latent, below=of alpha, label={[label distance=-3pt]250:{$\theta$}}] (theta) {} ;
  
  \node[latent, below=of theta, label=below:{$r$}] (r) {$|L_i|$} ;
  \node[obs, right=of r, label=below:{$f$}] (f) {} ;

  \edge[->] {alpha, beta} {theta} ;
  \edge[->] {theta} {r} ;
  
  \edge[->] {f} {r} ;
  
  \plate {buckets} {(theta)} {$|\mathcal{L}|$} ;
  \plate {cards} {(r)(f)} {$|\mathcal{C}|$} ;
  
%
%

\end{tikzpicture} 
  \caption{\textbf{Plate notation of the probabilistic graphical model.} The blames $b_{ij}$ are a direct function of $\bm{\theta}$ and $\mathbf{f}$ and are omitted for clarity.} 
  \label{fig:platemodel}
\end{figure}

This model can be formally defined as follows: 
\begin{align}
 \theta_j & \sim Beta(\alpha, \beta)\\
 b_{ij} & =
  \begin{cases}
   \frac{\theta_j}{\sum_{k \in N_i} \theta_k}        & \text{if } f_i = 1 \text{ and } j \in \mathbf{L_i} \\
   0        & \text{otherwise}
  \end{cases} \label{eq:poctoblames} \\
 r_i & \sim Categorical(|L_i|, \mathbf{b_i})
\end{align}
whose corresponding graphical model in plate notation can be seen in Figure \ref{fig:platemodel}.
The only hyper-parameter of this formulation is the Beta distribution which is encoded using $\alpha$ and $\beta$. Intuitively, $\alpha$ and $\beta$ control how much evidence we need to be confident that a location has really been compromised. 

The inputs of this model are the sets $\mathbf{L}_i$ (the locations with which each card interacted) and the boolean indicators $f_i$ on whether a card has been a victim.
Note the direct relationship between the problem definition and this formulation: the probability that a location $j$ has been compromised can be obtained directly from the expected value of $p_j$ ($E[p_j] = \theta_j$) and the blames attributed by the different cards are encoded in the row-normalized matrix $\mathbf{B}$.

In the following, we will describe an alternating algorithm to simultaneously find $\bm{\theta}$ and $\mathbf{B}$. 

\subsection{From Blames to POC Probabilities}
Let's suppose that we knew $\mathbf{B}$, i.e., we knew how much blame each card attributes to each possible POC. Ideally, we would then like to find $\bm{\theta}$ that maximizes $P[\bm{\theta} | \mathbf{B}]$, as our model relates the probability of being compromised with the blames attributed. 

We defined that $\theta_j$ comes from a $Beta(\alpha, \beta)$ distribution, therefore we know that:
\begin{equation}
 P[\theta_j ; \alpha, \beta] = \frac{\theta_j^{\alpha-1} (1-\theta_j)^{\beta-1}}{B(\alpha, \beta)},
\end{equation}
where the beta function $B(\alpha,\beta)$ is simply the normalization constant that ensures that the total probability integrates to 1.

Let $z_j = \sum_i b_{ij}$ be the sum of all the blames assigned to location $j$. $z_j$ follows a Beta-Binomial distribution and we know that 
\begin{equation}
 P[z_j | \theta_j ; |N_j|] \propto \theta_j^{z_j}(1-\theta_j)^{|N_j|-z_j}
\end{equation}
and, from Bayes' theorem, the posterior distribution equals
\begin{equation}
  \begin{gathered}
  P[\theta_j | z_j ; \alpha, \beta, |N_j|] \propto P[z_j | \theta_j ; |N_j|] P[\theta_j ; \alpha, \beta] \propto \\
  \propto \theta_j^{z_j+\alpha-1} (1-\theta_j)^{|N_j|+\beta-z_j-1}
  \end{gathered}
\end{equation}

This means that the posterior probability distribution of $\theta_j$ is defined as another Beta distribution that can be parametrized as $Beta(z_j+\alpha, |N_j|-z_j+\beta)$ and with expected value 
\begin{equation}
 E[\theta_j] = \frac{z_j+\alpha}{|N_j|+\alpha+\beta} \label{eq:expectedPOC}
\end{equation}

We can think of this expected value as a ratio of the blames ($z_j$) to the total number of cards that used this location ($|N_j|$), with additional terms $\alpha$ and $\beta$ that represent, respectively, ``virtual'' \textit{fraud-cards} that used this location ($\alpha$) and ``virtual'' non-\textit{fraud-cards} that used this location ($\beta$). Depending on the $Beta(\alpha, \beta)$ prior chosen, two possible POCs with the same fraud-cards ratio will have their probability adjusted to match our confidence on how far away from the prior distribution they are. The ratio $\displaystyle\frac{\alpha}{\alpha+\beta}$ represents the expected probability that a random location is compromised, while the magnitude of $\alpha$ and $\beta$ encode our confidence on this prior.


%
%
 
\subsection{From POC Probabilities to Blames}
Following a similar line of reasoning, let's suppose that we knew $\bm{\theta}$ and that we would like to find $\mathbf{B}$, i.e. we would like to know the probability that a card will blame each location, given their respective compromise probabilities.

As mentioned in Section \ref{sec:model}, we assume that blaming probabilities are linearly proportional to compromise probabilities. Therefore, the blames matrix $\mathbf{B}$ can be found by directly applying Equation \ref{eq:poctoblames}.

\subsection{An Alternating Algorithm}
We previously defined the probability that a location was compromised based on the blames assigned to it, and defined the blame assigned by a card to a possible POC given the compromise probabilities of all the locations in Equation \ref{eq:poctoblames}. This tight coupling suggests an alternating algorithm in which one updates blames and POC probabilities in succession until convergence. 
 
We initialize blames as uniformly distributed across all neighboring possible POCs, and proceed by updating the POC probabilities and blames in sequence. We check for convergence using the $l_1$ norm of the difference of successive POC probability estimations. Algorithm \ref{alg:poc} describes this procedure and, as we will see in Section \ref{sec:parallel}, one of its advantages is being easily parallelizable.
 
 \begin{algorithm}[ht]                      
\caption{\methodsc}          
\label{alg:poc}
\begin{algorithmic}
  \Require $\mathbf{L}$, locations neighboring of each card
  \Require $\mathbf{N}$: cards neighboring each location
  \Require $\epsilon$: convergence threshold
  \Require $\alpha, \beta$: prior parameters
  \\
  \State $\mathbf{B} \gets$ \textsc{UniformBlames}()
  \State $\bm{\theta} \gets$ \textsc{UpdatePOCProbabilities}($\mathbf{B}$, $\mathbf{N}$, $\alpha$, $\beta$)
  \Repeat
    \State $\mathbf{B} \gets$ \textsc{UpdateBlames}($\bm{\theta}$, $\mathbf{L}$)
    \State $\bm{\theta_{prev}} \gets \bm{\theta}$
    \State $\bm{\theta} \gets$ \textsc{UpdatePOCProbabilities}($\mathbf{B}$, $\mathbf{N}$, $\alpha$, $\beta$)
  \Until{$\Vert \bm{\theta} - \bm{\theta_{prev}}\Vert_1 < \epsilon$}
  \State\Return $(\bm{\theta},\mathbf{B})$
\\

\Function{UpdateBlames}{$\bm{\theta}$, $\mathbf{L}$}
  \For{every card $i$}
    \State $sum \gets \displaystyle\sum_{k \in L_i}\theta_k$
    \For{every location $j \in L_i$}
      \State $b_{ij} \gets \displaystyle\frac{\theta_j}{sum}$
    \EndFor
  \EndFor
  \State\Return $\mathbf{B}$
\EndFunction
\\

\Function{UpdatePOCProbabilities}{$\mathbf{B}$, $\mathbf{N}$, $\alpha$, $\beta$}
  \For{every location $j$}
    \State $z_j \gets \sum_{i} b_{ij}$
    \State $\theta_j \gets \frac{z_j + \alpha}{|N_j|+\alpha+\beta}$
  \EndFor
  \State\Return $\bm{\theta}$
\EndFunction
\end{algorithmic}
\end{algorithm}

\subsection{Convergence}
\begin{figure}[ht] 
  \centering 
  \includegraphics[width=0.45\textwidth]{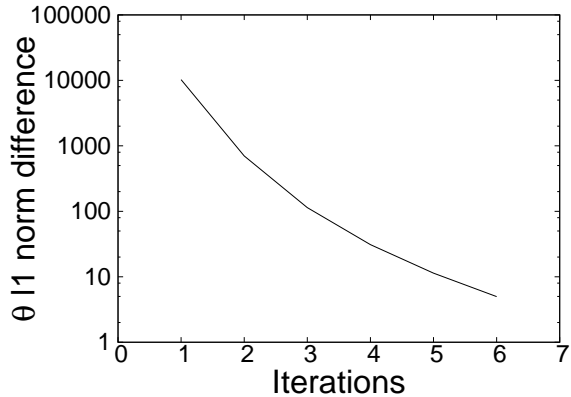} 
  \caption{\textbf{Exponentially fast convergence - } notice the log scale in the y-axis.} 
  \label{fig:convergence}
\end{figure}

Given its many applications, such as k-means clustering or expectation-maximization methods, the convergence of alternating optimization algorithms is a well studied problem and it is known to work well in a surprising number of cases \cite{bezdek2002alternating}. In general, one cannot guarantee global convergence but local convergence tends to be very fast. By using the dataset and hyper-parameters described in Section \ref{sec:results}, we show empirically that our implementation of Algorithm \ref{alg:poc} converges exponentially fast in Figure \ref{fig:convergence}.

%
%
 

\section{Distributed POC-detection} \label{sec:parallel}
Both stages of the alternating optimization algorithm are parallelizable if we assume a message-passing model of computation, as used in Pregel \cite{malewicz2010pregel} and other large-scale graph processing systems. We assume that both nodes and edges can store information which is shared and updated until the whole system converges; in our case, edges contain information relative to blames ($\mathbf{B}$) while location nodes contain their respective \textit{Point-of-Compromise} probability.

Under this new model of computation, the differences to Algorithm \ref{alg:poc} can be summarized in Algorithm \ref{alg:distpoc}.

\begin{algorithm}[ht]                      

\renewcommand{\algorithmicdo}{\textbf{do in parallel}}
\renewcommand{\algorithmicforall}{\textbf{for each}}

\caption{Distributed \methodsc}          
\label{alg:distpoc}
\begin{algorithmic}
\State G.POCs.onNewMessage($blame$)
\Indent
\State $z = z + blame$
\State $\Theta = \frac{z + \alpha}{|N_j| + \alpha + \beta}$
\EndIndent
\\

\State G.Cards.onNewMessage($\theta$)
\Indent
\State $sum = sum + \theta$
\EndIndent
\\

\Function{UpdateBlames}{$\bm{\theta}$, $\mathbf{N}$}
  \ForAll{location j in $\mathcal{G}$}
      \ForAll{Card i in $N_j$}
	  \State j.sendMessage(i, $\theta$)
      \EndFor
  \EndFor
  \State // After all messages are aggregated.
  \ForAll{Edge e in $\mathcal{G}$}
    \State e.blame = $\frac{e.POC.\theta}{e.card.sum}$
  \EndFor
\EndFunction
\\

\Function{UpdatePOCProbabilities}{$\mathcal{G}$}
  \ForAll{Edge e in $\mathcal{G}$}
    \State e.sendMessage(e.POC, e.blame)
  \EndFor
\EndFunction
\end{algorithmic}
\end{algorithm}

%
%
%

\textbf{Implementation.}
In order to obtain an highly efficient parallel solution, this implementation was developed using Apache Spark \cite{Spark}, a MapReduce engine that enables in-memory computation. Spark is well suited for machine learning algorithms as its in-memory model doesn't force sequential stages to synchronize data to disk. 

In particular, we rely on Spark's GraphX \cite{GraphX} module which overlays an abstraction for graph-parallel computation that allows message passing and aggregation.

\begin{table*}[t]
 \centering
 \caption{Several Points-of-Compromise identified in one of the datasets have also been mentioned in news reports.}
 \label{tab:realPOCs}
 \begin{tabular}{ | p{0.17\textwidth} | p{0.21\textwidth} | p{0.55\textwidth} |}
    \hline
    \textbf{Merchant} & \textbf{Source} & \textbf{Summary} \\ \hline
    Schnucks & \textit{ComputerWorld}~\cite{schnucksComputerWorld} & A supermarket chain where a breach exposed 2.4M cards. \\ \hline
    NoMoreRack.com & \textit{Reuters}~\cite{nomorerackReuters} & An online retailer with over \$340 million in sales annually, probes likely card breach. \\ \hline
    Jetro Cash \& Carry & \textit{DataBreaches}~\cite{jetroDataBreaches} & A data breach allowed attackers to access private data in cards used over a one month period in this chain. \\ \hline
    Bashah's Family of Stores & \textit{BankInfoSecurity}~\cite{bashasBankInfoSecurity} & A supermarket chain tied to the compromise of hundreds of cards. \\ \hline
    Buy.com & \textit{Yahoo Finance - Money Talks}~\cite{buyYahoo} & Hundreds of online shoppers reported fraudulent charges on their credit cards after making a purchase at this online marketplace. \\ \hline
  \end{tabular}
\end{table*}

\section{Results} \label{sec:results}
We answer the following questions:

\ben[{\bf Q}1.]
\item \textbf{Effectiveness - } How accurately and how early can we detect \textit{Points-of-Compromise} in reality? What are the trade-offs?
\item \textbf{Scalability - } How does our method scale in terms of the size of the network and in terms of the number of machines available?
\item \textbf{Fraud-cards precision and recall - } How much of the fraud that is reported can be explained through the identification of \textit{Points-of-Compromise}?
\item \textbf{Discoveries - } Can we identify real and validated \textit{Points-of-Compromise} in real data?
\een

\subsection{Experimental Setup}
\methodsc was applied to two datasets provided by different sources, each with over one billion transactions, 0.4 million cards and 2 million fraudulent transactions that cover more than one year. The data is very unbalanced with the percentage of transaction fraud in accordance with industry averages, between 0.01\% and 0.1\%~\cite{ecbcardfraud}. Due to privacy concerns, results in this section do not indicate the corresponding dataset.

\begin{table}[h]
 \centering
 \caption{Overview of the two datasets used. Specific values masked for privacy.}
  \begin{tabular}{|>{\hspace{-2pt}}l<{\hspace{-2pt}}|>{\hspace{-1pt}}c<{\hspace{-1pt}}|>{\hspace{-1pt}}c<{\hspace{-1pt}}|>{\hspace{-1pt}}c<{\hspace{-1pt}}|>{\hspace{-1pt}}c<{\hspace{-1pt}}|}
    \hline
    & \#cards & \#trxs & \#buckets & \#fraud trxs \\ \hline
    Dataset1 & $> 10^5$ & $> 10^9$  & $> 10^6$ & $> 10^6$ \\ \hline
    Dataset2 & $> 10^5$ & $> 10^9$  & $> 10^6$ & $> 10^6$ \\ \hline
  \end{tabular}
\end{table}

We created possible POCs corresponding to \textit{terminal-week} pairs and removed multi-edges and all possible POCs that interacted with less than 5 \textit{fraud-cards}, as they could not be confidently labeled \textit{Points-of-Compromise} under any circumstance. After this pre-processing stage, there were at least 1.5 million terminal-week pairs that had to be considered in each dataset. Results here reported correspond to a $\alpha = 0.2$ and $\beta = 15$ prior which provides a significant assumption that a random \textit{terminal-week} is not compromised. Results did not differ significantly with other values of $\alpha$ and $\beta$ we tested.


\subsection{Empirical Evidence and Fraud Prevented}\label{sec:empirical}
We collected significant empirical evidence demonstrating our ability to find real POCs, both data breaches and terminals that we suspect were victims of skimming, through manual analysis of the POCs reported. The list includes tobacco machines equipped with credit card readers and other general vending machines where the percentage of \textit{fraud-cards} is as high as 80\%.

Data breaches are easier to validate as they are often reported in the news; a non-exhaustive list of POCs we automatically detected along with a sample news report can be found in Table \ref{tab:realPOCs}. We were also able to identify POCs whose first news report occurred more than 6 months after the last transaction available in the dataset.

We evaluated the amount of fraud that could be prevented if cards of likely-compromised POCs were automatically reissued. Figure \ref{fig:gain} shows the gains obtained when \methodsc is evaluated in a 6 weeks period. Over \$2 million USD in savings would be possible when reissuing all cards that interacted with POCs with an expected compromise probability above 10\%, even if we assume a \$10 reissue cost per card. 17\% of the cards reissued would have been victims of fraud and 95\% of these would be first-time victims.

\subsection{Accuracy and Early Detection}\label{sec:synthetic}

\begin{figure*}[t]
    \centering
    \subfloat[Receiver Operating Curve: notice the \textbf{very low false positive rate}.]{\includegraphics[width=0.40\textwidth]{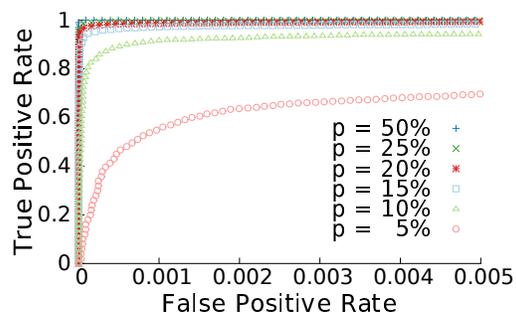} \label{fig:noise0xroc}}
    \hfil
    \subfloat[\textbf{High precision and recall, even with low stealing probability ($p$).}]{\includegraphics[width=0.40\textwidth]{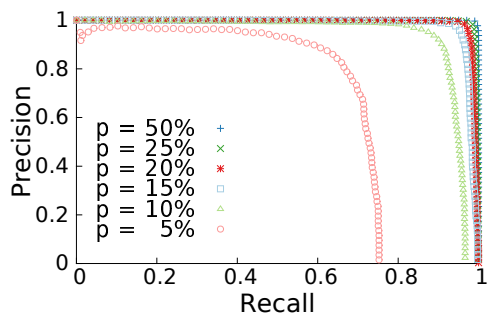} \label{fig:noise0xpvr}}
    
    \subfloat[Receiver Operating Curve: \textbf{low false positive rate, with 100\% noise}.]{\includegraphics[width=0.40\textwidth]{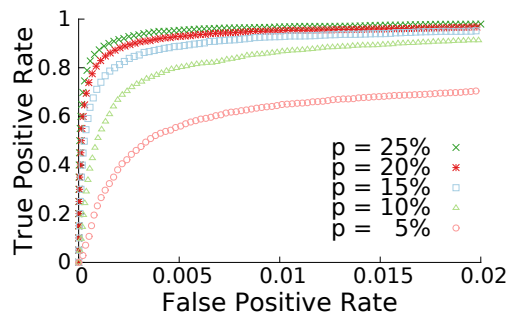} \label{fig:noise1xroc}}
    \hfil
    \subfloat[\textbf{Even with 100\% added noise, high precision and recall.}]{\includegraphics[width=0.40\textwidth]{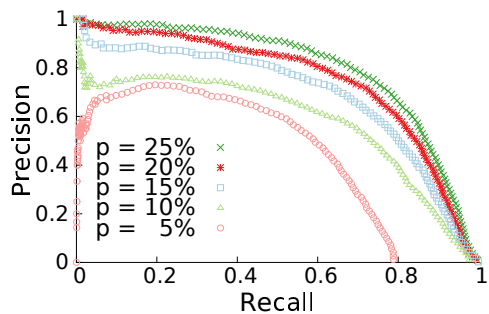} \label{fig:noise1xpvr}}

    \caption{Accuracy with varying probability of a card being a victim of fraud. (Top row: without noise. Bottom row: 100\% additional fraud-cards as noise.)}
    \label{fig:noise0x}
\end{figure*}

Due to the lack of ground-truth regarding which locations are effectively \textit{Points-of-Compromise}, we evaluate the precision and recall of \methodsc through the injection of synthetic POCs in real data. We evaluate \methodsc along two dimensions: 
\begin{enumerate}
 \item Probability that a card is a victim of fraud after using a compromised location ($p$). This can be viewed as a proxy for how early our method is able to detect compromised cards, as detection gets naturally easier as the number of victims increases.
 \item Presence of noise in the set of \textit{fraud-cards}, i.e., we randomly mark additional cards as victims, although fraud cannot be attributed to any of the POCs.
\end{enumerate}

In each experiment, we define a set of POCs and vary a probability (p) that their transactions will steal the corresponding card. Based on this new \textit{fraud-cards} list, we then obtain a new list of possible \textit{Points-of-Compromise}. Table \ref{tab:prob_vs_poc} shows the number of \textit{fraud-cards} and \textit{possible Points-of-Compromise} as the probability of the card being stolen increases, when no noise is included.

\begin{table}[ht]
  \centering
  \caption{Impact of the infection probability on the number of \textit{fraud-cards} and possible Points-of-Compromise.}
  \label{tab:prob_vs_poc}  
  \begin{tabular}{|c|c|c|}
    \hline
    \textbf{Probability ($p$)} & \textbf{\#fraud-cards} & \textbf{\#poss. POC} \\ \hline
    5\% & \numprint{29326} & \numprint{104185} \\ \hline
    10\% & \numprint{57593} & \numprint{263115} \\ \hline
    15\% & \numprint{84833} & \numprint{450609} \\ \hline
    20\% & \numprint{110955} & \numprint{662967} \\ \hline
    25\% & \numprint{136896} & \numprint{892119} \\ \hline
    50\% & \numprint{257609} & \numprint{2168733} \\
    \hline
  \end{tabular}
\end{table}


Figures \ref{fig:noise0xroc} and \ref{fig:noise0xpvr} show the receiver operating characteristic (ROC) and ``precision vs recall'' curves for the different compromise probabilities. Note that we are able to simultaneously achieve high precision and recall for relatively small compromise probabilities; we achieve over 90\% precision and recall even when only 10\% of the cards have been victims of fraud. 

We also analyze the impact of noise in the effectiveness of our method: we double the number of \textit{fraud-cards} by randomly selecting additional cards. These are cards that do not have a corresponding POC in the data, even though they were marked as victims. As before, Figures \ref{fig:noise1xroc} and \ref{fig:noise1xpvr} show the ROC and ``precision vs recall'' curves for the different compromise probabilities. Using the same scenario of a 10\% probability of cards being compromised as example, note that we are still able to achieve about 75\% recall maintaining 50\% precision, even though we are simultaneously considering very aggressive levels of cards mislabeled as \textit{fraud-cards} and early detection.

\subsection{Scalability}
\begin{figure}[ht] 
  \centering 
  \includegraphics[width=0.39\textwidth,trim=0mm 1mm 0mm 1mm, clip]{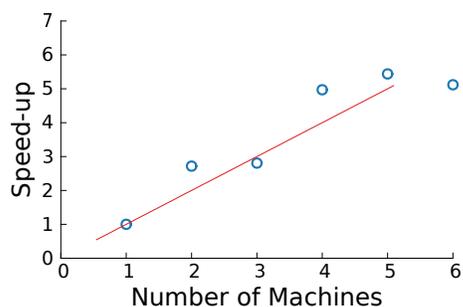} 
  \caption{\textbf{Good distributed speed-up.}} 
  \label{fig:cores_speedup}
\end{figure}

Scalability experiments are performed using the data described in Table \ref{tab:prob_vs_poc}. We show \methodsc linear scalability on the number of possible POCs (Figure \ref{fig:edges_time}) and analyze the runtime of its Spark implementation when changing the number of machines available in a small cluster of 6 quad-core machines (Figure \ref{fig:cores_speedup}). The number of cores in each machine does not provide any advantage, as disk input-output is the bottleneck of our system, not processing power.

%
%
%

\subsection{Comparison}\label{sec:comparison}
We compare the precision and recall of our method to (1) \textsc{FaBP} - Fast Belief Propagation~\cite{koutra2011unifying}; (2) a greedy approximation of Vertex Cover described in Section \ref{sec:backrw}; (3) the ratio as proxy for POC probability, as commonly used by previous methods and (4) the ratio metric combined with the best prior found for \methodsc, in order to reduce its false positive rate. 

\textsc{FaBP}~\cite{koutra2011unifying} is a fast approximation to Belief Propagation with low sensitivity to input parameters. We assigned a high prior belief (0.5) to \textit{fraud-card} nodes, a low prior belief (-0.1) to non-\textit{fraud-card} nodes and a neutral belief (0.0) to possible POC nodes. We then decreasingly sort the possible POCs by their final belief, creating the corresponding precision vs recall curve.
The curve for the other methods was created similarly, based on the ordering of the possible POCs that they explicitly define. 

We compare all methods on the non-noise dataset when considering a stealing probability of 10\%.

As can be seen in Figure \ref{fig:fabppvr}, our method significantly improves over all alternatives. Reasons have been detailed in previous sections, but can be summarized as a combination of appropriate priors and the focused blame of the \textit{fraud-cards}. Vertex Cover's result shows that focused blames are not sufficient, while the ratio with prior's result shows that removing false positives with a small amount of \textit{fraud-cards} is not enough either.

\subsection{Reproducibility}
We make available the dataset used in the comparison experiments described in Section \ref{sec:synthetic}, when requested by email to the authors.

\section{Conclusion}
We present, as far as the authors know, the first distributed procedure for the automatic detection of \textit{Points-of-Compromise} in transaction networks. We achieve highly accurate results through the implementation of an in-memory algorithm that updates POC probabilities and blame scores alternatingly, and we have demonstrated surprising empirical evidence in a real dataset. Our main contributions are the following:

\ben
 \item \textbf{Point-of-Compromise Problem.} We formulate a novel and important POC detection problem. 
 \item \textbf{Effectiveness.} \methodsc accurately identifies \textit{Points-of-Compromise}, achieving over 90\% precision and recall when only 10\% of the stolen cards have been used in fraud (Figure \ref{fig:fabppvr}).
 \item \textbf{Distributed POC-Detection algorithm.} We provide a scalable distributed algorithm for POC detection in big datasets (Figure \ref{fig:edges_time}).
\een

%

\section{Acknowledgments}
Miguel Araujo is partially supported by the FCT through the CMU$\vert$Portugal Program under Grant SFRH/BD/52362/2013.

\vspace{-1mm}

\bibliographystyle{abbrv}
\bibliography{icdm16}

\begin{thebibliography}{10}

\bibitem{jetroDataBreaches}
Restaurant depot/jetro cash \& carry customers’ credit cards hacked.
\newblock DataBreaches.net, December 2011.

\bibitem{buyYahoo}
B.~Ballenger.
\newblock Rakuten.com customers reporting credit card fraud.
\newblock finance.yahoo.com, June 2013.

\bibitem{bezdek2002alternating}
J.~C. Bezdek and R.~J. Hathaway.
\newblock Some notes on alternating optimization.
\newblock In {\em Advances in Soft Computing—AFSS 2002}, pages 288--300.
  Springer, 2002.

\bibitem{bolton2002statistical}
R.~J. Bolton and D.~J. Hand.
\newblock Statistical fraud detection: A review.
\newblock {\em Statistical science}, pages 235--249, 2002.

\bibitem{dal2014learned}
A.~Dal~Pozzolo, O.~Caelen, Y.-A. Le~Borgne, S.~Waterschoot, and G.~Bontempi.
\newblock Learned lessons in credit card fraud detection from a practitioner
  perspective.
\newblock {\em Expert systems with applications}, 41(10):4915--4928, 2014.

\bibitem{ecbcardfraud}
{European Central Bank}.
\newblock Fourth report on card fraud.
\newblock Technical report, European Central Bank, 2015.

\bibitem{fico2015}
{\relax Fair Isaac Corporation}.
\newblock How europe's card fraud is evolving.
\newblock Insights White Paper, 2015.

\bibitem{patentUS20050055373A1}
G.~Forman.
\newblock Determining point-of-compromise.
\newblock US Patent US 20050055373 A1, March 2005.

\bibitem{karp1972reducibility}
R.~M. Karp.
\newblock {\em Reducibility among combinatorial problems}.
\newblock Springer, 1972.

\bibitem{bashasBankInfoSecurity}
T.~Kitten.
\newblock Bashas' breach exposes security flaws.
\newblock BankInfoSecurity.com - The Fraud Blog, February 2013.

\bibitem{patentUS7580891B2}
V.~F. Klebanoff.
\newblock Method and system for assisting in the identification of merchants at
  which payment accounts have been compromised.
\newblock US Patent US 8473415 B2, August 2009.

\bibitem{koutra2011unifying}
D.~Koutra, T.-Y. Ke, U.~Kang, D.~H.~P. Chau, H.-K.~K. Pao, and C.~Faloutsos.
\newblock Unifying guilt-by-association approaches: Theorems and fast
  algorithms.
\newblock In {\em Machine Learning and Knowledge Discovery in Databases}, pages
  245--260. Springer, 2011.

\bibitem{targetKrebs}
B.~Krebs.
\newblock The target breach, by the numbers.
\newblock KrebsOnSecurity.com, May 2014.

\bibitem{lexisnexis201509}
{\relax LexisNexis Risk Solutions}.
\newblock Merchants contend with increasing fraud losses as remote channels
  prove especially challenging.
\newblock LexisNexis True Cost of Fraud Study, September 2015.

\bibitem{malewicz2010pregel}
G.~Malewicz, M.~H. Austern, A.~J. Bik, J.~C. Dehnert, I.~Horn, N.~Leiser, and
  G.~Czajkowski.
\newblock Pregel: a system for large-scale graph processing.
\newblock In {\em Proceedings of the 2010 ACM SIGMOD International Conference
  on Management of data}, pages 135--146. ACM, 2010.

\bibitem{ftpartners2015}
\relax Financial Technology Partners~Research.

\bibitem{nomorerackReuters}
A.~Roy.
\newblock Online retailer nomorerack.com probes likely card breach-report.
\newblock Reuters, March 2013.

\bibitem{patentUS8473415B2}
K.~P. Siegel, R.~A. Paynter, R.~L. Grossman, C.~Brown, C.~R. Byce, T.~Dwyer,
  and A.~Chen.
\newblock System and method for identifying a point of compromise in a payment
  transaction processing system.
\newblock US Patent US 8473415 B2, 2013 June.

\bibitem{schnucksComputerWorld}
J.~Vijayan.
\newblock Schnucks supermarket chain struggled to find breach that exposed 2.4m
  cards.
\newblock ComputerWorld, April 2015.

\bibitem{GraphX}
R.~S. Xin, J.~E. Gonzalez, M.~J. Franklin, and I.~Stoica.
\newblock Graphx: A resilient distributed graph system on spark.
\newblock In {\em First International Workshop on Graph Data Management
  Experiences and Systems}, GRADES '13, pages 2:1--2:6, New York, NY, USA,
  2013. ACM.

\bibitem{yamaguchi2015socnl}
Y.~Yamaguchi, C.~Faloutsos, and H.~Kitagawa.
\newblock Socnl: Bayesian label propagation with confidence.
\newblock In {\em Advances in Knowledge Discovery and Data Mining}, pages
  633--645. Springer, 2015.

\bibitem{patentUS8600872B1}
S.~Yan.
\newblock System and method for detecting account compromises.
\newblock US Patent US 8600872 B1, December 2013.

\bibitem{Spark}
M.~Zaharia, M.~Chowdhury, M.~J. Franklin, S.~Shenker, and I.~Stoica.
\newblock Spark: Cluster computing with working sets.
\newblock In {\em Proceedings of the 2nd USENIX Conference on Hot Topics in
  Cloud Computing}, HotCloud'10, pages 10--10, Berkeley, CA, USA, 2010. USENIX
  Association.

\bibitem{patentUS7761379B2}
S.~M. Zoldi, L.~Wang, L.~Sun, and S.~G. Wu.
\newblock Mass compromise/point of compromise analytic detection and
  compromised card portfolio management system.
\newblock US Patent 7761379 B2, July 2010.

\end{thebibliography}

\end{document}